\documentclass[conference]{IEEEtran}
\ifCLASSINFOpdf
\else
\fi
\usepackage{graphics} 
\usepackage{epsfig} 
\usepackage{amsmath} 
\usepackage{lipsum}
\usepackage{graphicx}
\usepackage{array}
\usepackage{algorithm}
\usepackage{amssymb}
\usepackage{balance}
\usepackage[noend]{algpseudocode}

\begin{document}
%
\title{Robust Power Scheduling for Microgrids with Uncertainty in Renewable Energy Generation}

\author{\IEEEauthorblockN{Amir Valibeygi, Abdulelah H. Habib, Raymond A. de Callafon}
\IEEEauthorblockA{Mechanical and Aerospace Engineering\\
University of California San Diego, La Jolla, California\\
avalibey@eng.ucsd.edu}
}


%


\maketitle

\begin{abstract}
  A robust power scheduling algorithm is proposed to schedule power flow between the main electricity grid and a microgird with solar energy generation and battery energy storage subject to uncertainty in solar energy production. To avoid over-conservatism in power scheduling while guaranteeing robustness against uncertainties, time-varying "soft" constraints on the State of Charge (SoC) of the battery are proposed. These soft constraints allow SoC limit violation at steps far from the current step but aim to minimize such violations in a controlled manner. The model predictive formulation of the problem over a receding time horizon ensures that the resulting solution eventually conforms to the hard SoC limits of the system at every step. The optimization problem for each step is formulated as a quadratic programming problem that is solved iteratively to find the soft constraints that are closest to the hard ones and still yield a feasible solution. Optimization results demonstrate the effectiveness of the approach.
%
%
\end{abstract}

\section{INTRODUCTION}

With the rapid growth of renewable Distributed Energy Resources (DERs) in power systems and due to their inherent uncertainties, economic energy management and planning is of great importance for microgrids of different scales. A Microgrid is defined as a group of interconnected DERs, loads, and storage units that act as a unified entity in the electricity market and is able to operate in both connected and islanded modes from the main electricity grid. In the connected mode, microgrids are connected to the main grid at the Point of Common Coupling (PCC) and any power exchange between the microgrid and the main grid is measured at this point. DERs within the microgrid may provide part or all of the microgrid's energy demand; hence reducing the energy drawn from the main grid. Electricity providers use different pricing schemes to encourage certain consumption behaviors among consumers and make power grids more efficient and reliable. Microgrids may use such energy price data to optimally schedule their loads, storage units, and dispatchable DERs \cite{shariatzadeh2015demand}.

The existence of storage provides additional flexibility to benefit from such pricing schemes. Optimal microgrid scheduling should take into account microgrid operational costs and seek to compute optimal storage/DER dispatch schedule \cite{parhizi2015market,habib2016model,khodaei2015microgrid}. This optimal power value should be computed based on load requirements, hardware constraints, storage capacity, and with the consideration of intermittent and uncertain nature of renewables. Load and renewable generation uncertainty are two major sources of uncertainty in microgrids that could highly impact its economic performance \cite{baziar2013considering}. Various approaches have been proposed to address uncertainties in the predicted load \cite{sarantis2017optimal, samadi2013tackling}. Renewable uncertainties generally impose a greater impact if a significant portion of microgrid energy is provided by renewable DERs. Various robust and stochastic microgrid scheduling methods have been studied to address renewable uncertainty challenges.

A widely popular approach to handle uncertainty is to formulate the problem as an stochastic programming problem by considering different scenarios for the uncertain parameters and their probabilities \cite{parisio2013stochastic}. The goal of the problem will then be minimizing cost of current decisions plus the expected value of cost of future decisions. In an alternative approach, uncertain generation can be handled by robust optimization formulations where all possible scenarios within the uncertainty set are accounted for and constraint satisfaction is guaranteed regardless of the realization of the random variables within a certain set \cite{li2011comparative,malysz2014optimal}. If a feasible solution exists for this problem, it tends to be more conservative yet less computationally intensive compared to the stochastic formulation. In a different approach, the problem can be studied within the framework of Chance Constrained Programming (CCP). Assuming known distribution for the uncertain variables, CCP can be used to compute a minimizing solution that meets inequality constraints with a certain probability \cite{wu2011economic,charnes1959chance}. Although less conservative for most realizations of the random variable, this approach may lead to constraint violation if knowledge of distribution is inaccurate or if extreme realizations of the random variables occur.

In the context of day-ahead microgrid scheduling, the intended scheduling interval is relatively large. In addition, forecast of uncertain renewable generation will be updated and more accurate forecasts will become available as time progresses. These are two motivations behind taking a model predictive approach to the scheduling problem. If such approach is taken, one need not determine a unique scheduling at the beginning of the 24-hr interval that is robust against all possible uncertainties. Instead, some possible realizations may be allowed to violate the conditions at steps far into the future. Future updates of the model predictive solution will guarantee that no actual condition violation will happen. This framework will however allow for more optimal (less conservative) solutions to the problem. Diferent works have taken a model predictive approach to the scheduling problem \cite{jin2017user,prodan2014model}. Reference \cite{parisio2013stochastic} compares the performance of stochastic and deterministic MPC in economic scheduling of microgrids. Also, \cite{zhang2016model} provides a model predictive economic scheduling solution and discusses the impact of forecast error.

In this work, microgrid optimal scheduling problem is considered for a grid-connected microgrid with solar generation and energy storage. Approximate solar forecast and its uncertainty bounds are assumed to be known for the day-ahead microgrid operation and are updated at 15-min intervals. To reduce conservatism in the presence of inherent uncertainties of PV generation, the problem is formulated within a model predictive framework and hard constraints are replaced with their relaxed versions at each step of MPC. 
At each step with the newly updated predictions, a quadratic programming problem is solved to yield an economic solution.

\textbf{Notation.}
Set of time intervals over 24 hours is denoted as $\mathcal{T}=\{1,2,...,T\}$. 
$s(t)$ is a random variable representing solar generation during interval $t$, while $\bar{s}(t,\tau),s_u(t,\tau),s_l(t,\tau)$ are solar generation forecast, its upper bound, and its lower bound during interval $t$ for predictions made at step $\tau$. In a similar way, $c(t)$ is a random variable showing the charge level of the battery during interval $t$ and $\bar{c}(t,\tau),c_u(t,\tau),c_l(t,\tau)$ are charge level forecast, its upper bound, and its lower bound during interval $t$ for MPC step $\tau$. In the remainder of the paper, $\tau$ indices for MPC steps are dropped for simplicity when it does not lead to confusion. $d(t), e(t),$ and $r(t)$ show load demand, energy to/from the inverter, and energy flow at PCC during interval $t$. Energy variables ($s,d,e,r$) are assumed uniform over the intervals and represent total energy over the 15-min interval.

\section{SYSTEM DESCRIPTION AND PROBLEM FORMULATION}

We study a grid-connected microgrid consisting solar energy generation, battery energy storage and several loads. Strict load requirements of the microgrid enforce that complete demand must be met at each time. The storage unit can be exploited to implement power shifting by using the stored energy at times of high market price and therefore reduce energy input from the main grid at those times. The economic scheduling problem is expected to propose a microgrid power flow solution at the point of common coupling that while reducing a certain cost function, meets microgird's operational constraints and guarantees robustness against PV prediction uncertainties. This work aims to develop a model-predictive robust optimization scheme based on 24-hr prediction of solar energy, load demand, and price. The 24-hr horizon is divided into $\Delta_T=15$~min intervals and the optimization algorithm for the remaining intervals of the 24-hr period is run at the beginning of each MPC step. However, only the results for the impending interval will be implemented. We first formulate the benchmark scheduling problem without uncertainties.
\subsection{Benchmark Problem}
The microgrid is assumed to have a certain energy demand for each of the considered intervals $t \in \mathcal{T}$. The sum of energy from the main grid and energy from DER/storage during each interval should meet this load demand
\begin{align}
d(t) = e(t) + r(t)
\label{demand_balance}
\end{align}
Additionally, variation in the charge level of the battery can be expressed as
\begin{align}
c(t+1) = c(t) + s(t) - e(t)
\label{battery_balance}
\end{align}
It should be noted that as indicated in this equation, a unique decision variable $e(t)$ (inverter energy to/from microgrid) should be applicable regardless of the actual realization of PV generation. This means that uncertainty in PV generation should only translate into uncertainty in the state of charge of the battery.

\textbf{Constraints.} The microgrid optimal scheduling problem should be solved with the consideration of operational constraints of the system. A few of the most essential constraints are formulated in this section. The inverter can only provide power to the microgrid subject to its operational limits
\begin{align}
P_{min}.\Delta_T\leq e(t) \leq P_{max}.\Delta_T 
\label{inv_limits}
\end{align}
where $P_{min/max}$ is min/max inverter power.
The state of charge of the battery should stay within its upper and lower limits
\begin{eqnarray}
C_{min}<\bar{c}(t,\tau)<C_{max}
\label{SoC_constraint1}
\end{eqnarray}
The scheduling should further enforce that the final charge level of the battery at the end of the 24-hr interval is equal to its initial charge.
\begin{align}
\bar{c}(t=T,\tau)=C_0
\label{C_0}
\end{align}
Taking into account the last two constraints, the set of feasible battery charge levels can be denoted as
\begin{align}\nonumber
\mathcal{C}=\{c(t,\tau) \mid C_{min}&<c(t,\tau)<C_{max}\ \forall t \geq \tau;\\
& c(t=T,\tau)=C_0\}
\label{charge_constraint}
\end{align}

\textbf{Cost Function.} The optimal scheduling problem is expected to minimize a cost function comprising cost of electricity and other operational costs of the microgrid. Cost of energy provided by the renewable source is assumed to be negligible. The total cost associated with microgrid operation is formulated below. The first term accounts for energy charge during the 24-hr period and the second term shows demand charge incurred during the time interval with maximum energy consumption. The last cost term is a penalty on battery charge/discharge to reduce energy loss due to battery round-trip efficiency.
\begin{align}
f = \sum\limits_{t=1}^T v(t)r(t) + k r_{max} +  \alpha \sum\limits_{t=1}^T e(t)^2
\label{CF}
\end{align}
where $v(t)$ is electricity unit price at time $t$, $r_{max}$ is energy flow at PCC during the interval with maximum energy consumption, and $k$ and $\alpha$ are weighting coefficients.
\subsection{Renewable Forecast Uncertainty}
Numerous solar forecasting methods have been explored in the literature. In this work, we assume knowledge of expected value of solar generation during interval $t$ at MPC step $\tau$, $E\{s(t)\}_\tau = \bar{s}(t,\tau)$ as well as its upper and lower bounds $s_u(t,\tau),s_l(t,\tau)$ is available for all $t\geq \tau$. These bounds should guarantee with a high degree of confidence that the realized amount of solar generation will be within their range. This solar forecast data is updated with the most recent data at each step. Clearly, more accurate predictions and smaller uncertainty bounds are available if forecast interval $t$ is closer to current step $\tau$. For each MPC step $\tau$, we define the sequence $S_u:\{s_u(\tau),s_u(2), ...,s_u(T)\}$ as a solar scenario consisting of upper forecast points for all times $\tau \leq t \leq T$. This scenario represents an unlikely realization of $s(t)$ over the remainder of the 24-hr period which takes the upper solar forecast value at each time. In a similar way we define $S_l$ and $\bar{S}$ as the lower and average solar scenarios. 
In order to be able to implement the unique decision variable $e(t)$ for all possible solar scenarios, the uncertainty in solar generation should only translate into uncertainty in the state of charge of the battery. With the motivation to obtain a single scheduling plan that accommodates various solar generation scenarios $(\bar{S},S_u,S_l)$, we intend to extend the aforementioned problem into a robust scheduling problem.
\section{Uncertainty Handling and Robust Scheduling}
\subsection{Existing Approaches} A rudimentary approach would be to enforce the above hard constraint \eqref{SoC_constraint1} on charge level of all possible solar generation scenarios.
\begin{align}\nonumber
C_{min}<c_u(t,\tau)<C_{max}\\
C_{min}<c_l(t,\tau)<C_{max}
\end{align}
Such constraint over the 24-hr period could make the solutions highly conservative or even infeasible \cite{li2011comparative}. To reduce the conservatism knowing that the solution will be updated at later steps, a chance constrained formulation would require
\[
P(C_{min}<c_u(t,\tau)<C_{max}, C_{min}<c_l(t,\tau)<C_{max}]\geq \beta
\]
where $P(.)$ is probability of the constraint satisfaction.
This constraint is a rational relaxation of the previous constraint. It can therefore yield feasible/more optimal solutions for the next MPC step and update its solution at every step. It however requires knowledge of distribution of the uncertainty and demands higher computational complexity than the previous approach. A different approach would be to implement an additional cost term in the cost function associated with SoC limit violation in the place of SoC constraints for upper and lower SoC scenarios \cite{sarantis2017optimal}.
\begin{equation}
\begin{array}{c}
f_{aug} =  \sum\limits_{t=1}^T max\{0,c_u(t)-C_{max}\} \\ + \sum\limits_{t=1}^T max\{0,C_{min}-c_l(t)\}
\end{array}
\label{CF1}
\end{equation}
Although this additional cost term could act as a soft constraint on SoC limit violation, it has no structure to differentiate between intervals with and without uncertainty.
\subsection{Proposed Method}
\begin{figure}[thpb]
    \centering
    \includegraphics[width=0.75\columnwidth]{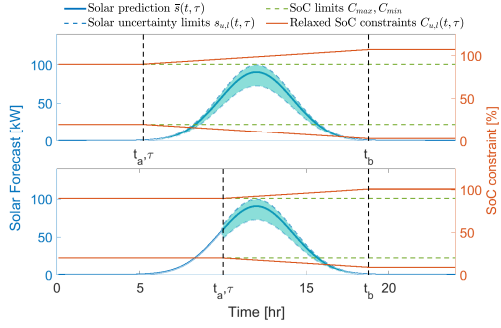}
    \caption{ \textbf{Top.} Uncertainty plot at the beginning of the 24-hr interval. Expected solar generation and its upper and lower uncertainty bounds are shown in blue. Hard SoC limits (green) are replaced by soft constraints (orange) to reach feasible solutions and reduce conservatism when there is uncertainty in solar prediction. \textbf{Bottom.}  Uncertainty plot at time $\tau=10~hr$ of the 24-hr interval. Updated solar forecast and its upper and lower uncertainty bounds for the remainder of the 24-hr interval $(t\geq \tau)$ are shown in blue.}
    \label{uncertainty1}
\end{figure}

We propose a framework to generalize the regular SoC constraints (\ref{charge_constraint}) in a structured way that makes it applicable to uncertain solar scenarios. The idea is to enforce the hard constraint (\ref{charge_constraint}) only on the expected value of solar forecast  and a relaxed version of it on other possible solar generation scenarios $(S_u,S_l)$. The result is that instead of limiting all possible charge level realizations to fall within upper and lower limits, we allow them to linearly grow out of bounds during uncertain intervals, but control their growth by a parameter representing tightness of the constraint. The rationale behind this scheme is that the uncertainty in accumulative solar generation is additive over time and therefore storage scheduling over time should allow more relaxed constraints in the future in order for the solution to remain feasible. We define parameter $\eta$ to characterize such relaxed constraints on SoC of battery under upper and lower PV scenarios. Figure \ref{uncertainty1}-top shows the proposed constraint on the battery charge level across different uncertainty scenarios at the beginning of the 24-hr interval. Also, Figure \ref{uncertainty1}-bottom shows the updated constraints at time $\tau$ for the remainder of the 24-hr interval. These constraints can be formulated as
\begin{eqnarray}\nonumber
C_{l}(t,\tau)<c_u(t,\tau)<C_{u}(t,\tau)\\
C_{l}(t,\tau)<c_l(t,\tau)<C_{u}(t,\tau)
\label{SoC_constraint2}
\end{eqnarray}
\begin{gather*}
C_{l[u]}(t,\tau)=
\begin{cases}
C_{min[max]}&  \hspace{8mm} t <t_a \ \\
C_{min[max]} -[+] (t-\tau).\eta & t_a<t<t_b \ \\
C_{min[max]} -[+] (t_b-\tau).\eta&  t_b<t\
\end{cases} 
\end{gather*}%
$t_a$ and $t_b$ are the time stamps of start and end of uncertainty in solar prediction. For $\eta=0$, this condition is equivalent to the strict conditions in (\ref{SoC_constraint1}). Applying the hard constraint ($\eta=0$) may yield no feasible solution over the entire 24-hr interval which means the problem was not solvable if hard SoC constraints were to be enforced on all possible scenarios at all times. For larger values of $\eta$, condition (\ref{SoC_constraint2}) means relaxation of conditions (\ref{SoC_constraint1}) as time progresses. Condition \eqref{SoC_constraint2} could be tested iteratively with different values of $\eta $ to obtain the smallest $\eta ~(\eta^*)$ for which a solution exists. Solutions will then exist for all $\eta>\eta^*$. The bigger the choice of $\eta$, the less conservative and less robust the solution will be.
The benefit of the alternative soft SoC constraint, as opposed to hard SoC requirement \eqref{SoC_constraint1}, is the reduced conservatism on the optimal solution freedom while keeping different scenarios under control. However, by running this algorithm at regular 15-min periods with updated forecast, we ensure that the resulting schedule will strictly meet hard SoC requirements \eqref{SoC_constraint1}.

\textbf{Robustness Analysis.} At each step of MPC, we want to make sure that the output $e(t)$ computed for the next step with soft SoC constraints does not lead to SoC limit violation. To achieve this, one can shift the soft constraint one step to the right so that the impending step always follows the hard limits while steps after that follow soft limits. Another less conservative approach would be to limit next step's optimal solution according to the following
\begin{align}
c(\tau)+s_u(\tau)-C_{max}\leq e(\tau)\leq c(\tau)+s_l(\tau)-C_{mn}
\end{align}

Based on the previous discussion, we can formalize Algorithm 1 for microgrid robust optimal scheduling.
\begin{algorithm}
\caption{Microgrid Robust Optimal Scheduling}\label{algorithm}
\begin{algorithmic}[1]
\State Start at the beginning of the 24-hr interval $\tau=1$
\State Obtain price and load demand data for the next 24-hr
\While{$\tau \leq T$}
\State Update solar forecast and its bounds for $t=\tau : T$
\State Initialize parameter $\eta = \eta_0$
\Repeat
\State Solve the quadratic programming problem with 

\quad cost function $(7)$ and subject to constraints

\quad $(1-5,11,12)$ using any available solver
\State Update $\eta$ using bisection to find smaller values

\quad of $\eta$ that gives a feasible solution
\Until{$|\eta_{current}-\eta_{last}|<\epsilon$}
\State $\eta^* = \eta$
\State Obtain Optimal scheduling solution with $\eta = \eta^*$ for

$t=\tau:T$ but implement only $e(t=\tau)$
\State Wait until the beginning of the next scheduling

period and arrival of updated forecast data
\EndWhile
\State \textbf{End while}
\end{algorithmic}
\end{algorithm}

\section{RESULTS}
\begin{figure}[thpb]
    \centering
        \includegraphics[width=6.5cm]{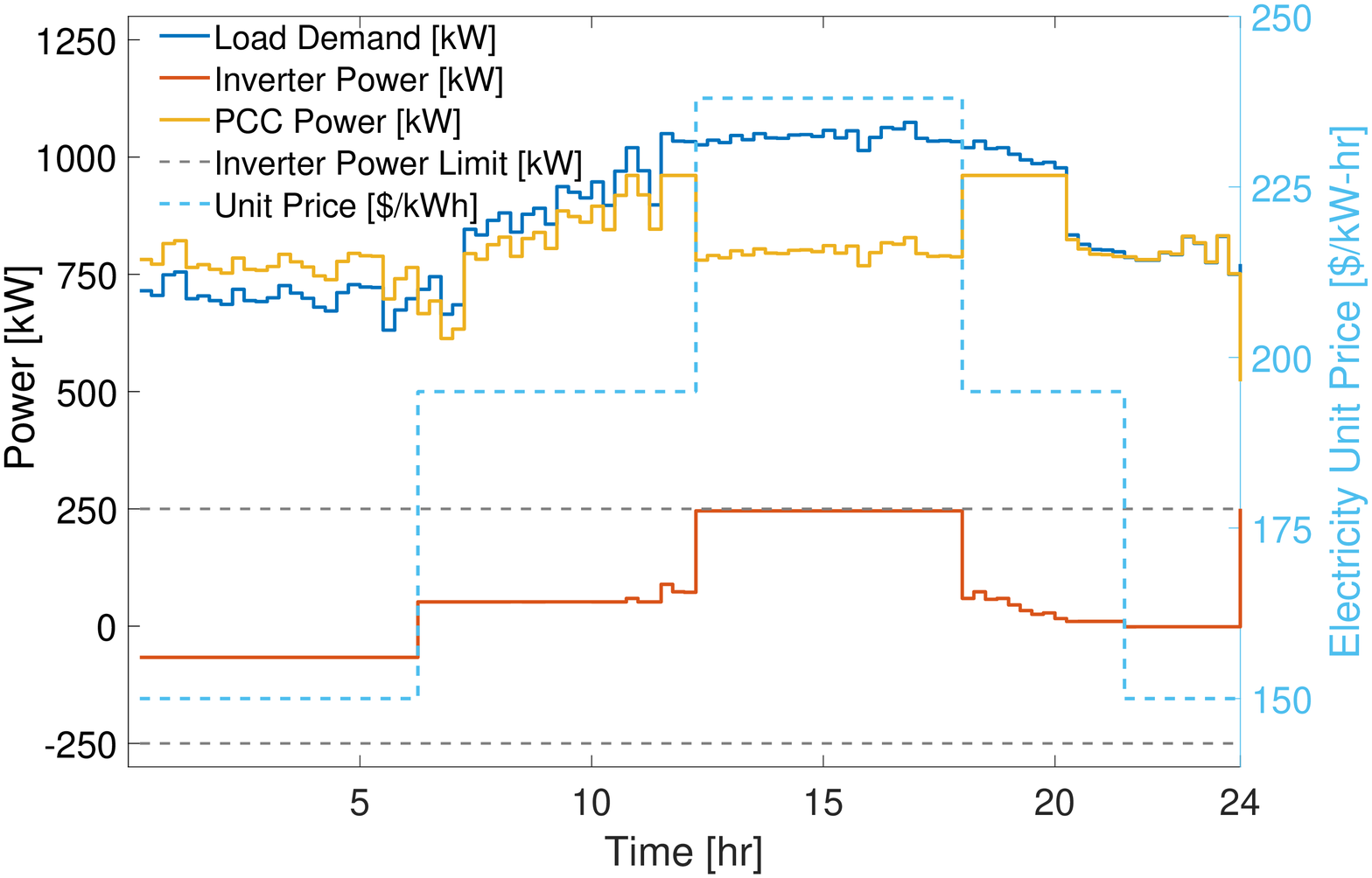}

    \caption{Optimization results for scheduling performed at time $\tau=1$.}
    \label{Power}
\end{figure}

We study the microgrid of a medical facility that is planning integration into the main grid. Solar generation forecast at the beginning of the 24-hr horizon and its uncertainty bounds for the considered facility are illustrated in Figure \ref{uncertainty1}.Load demand and price data are also illustrated in Figure \ref{Power}. Also, microgrid specifications are listed in table 1.

\begin{figure}[thpb]
    \centering
    \includegraphics[width=6.5cm]{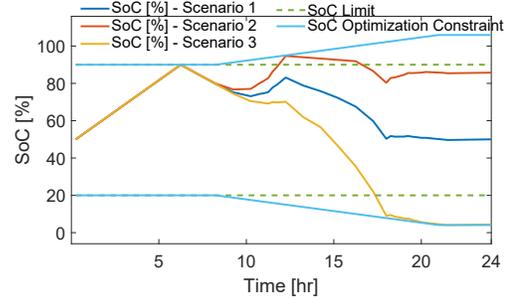}
    \caption{SoC variation within relaxed constraints for MPC step 1 (time step $\tau=1$) with expected $\bar{S}$, upper $S_u$, and lower $S_l$ solar generation scenarios}
    \label{SoC}
\end{figure}

\begin{figure}[thpb]
    \centering
    \includegraphics[trim=.8in 0 .5in 0,clip,width=\columnwidth]{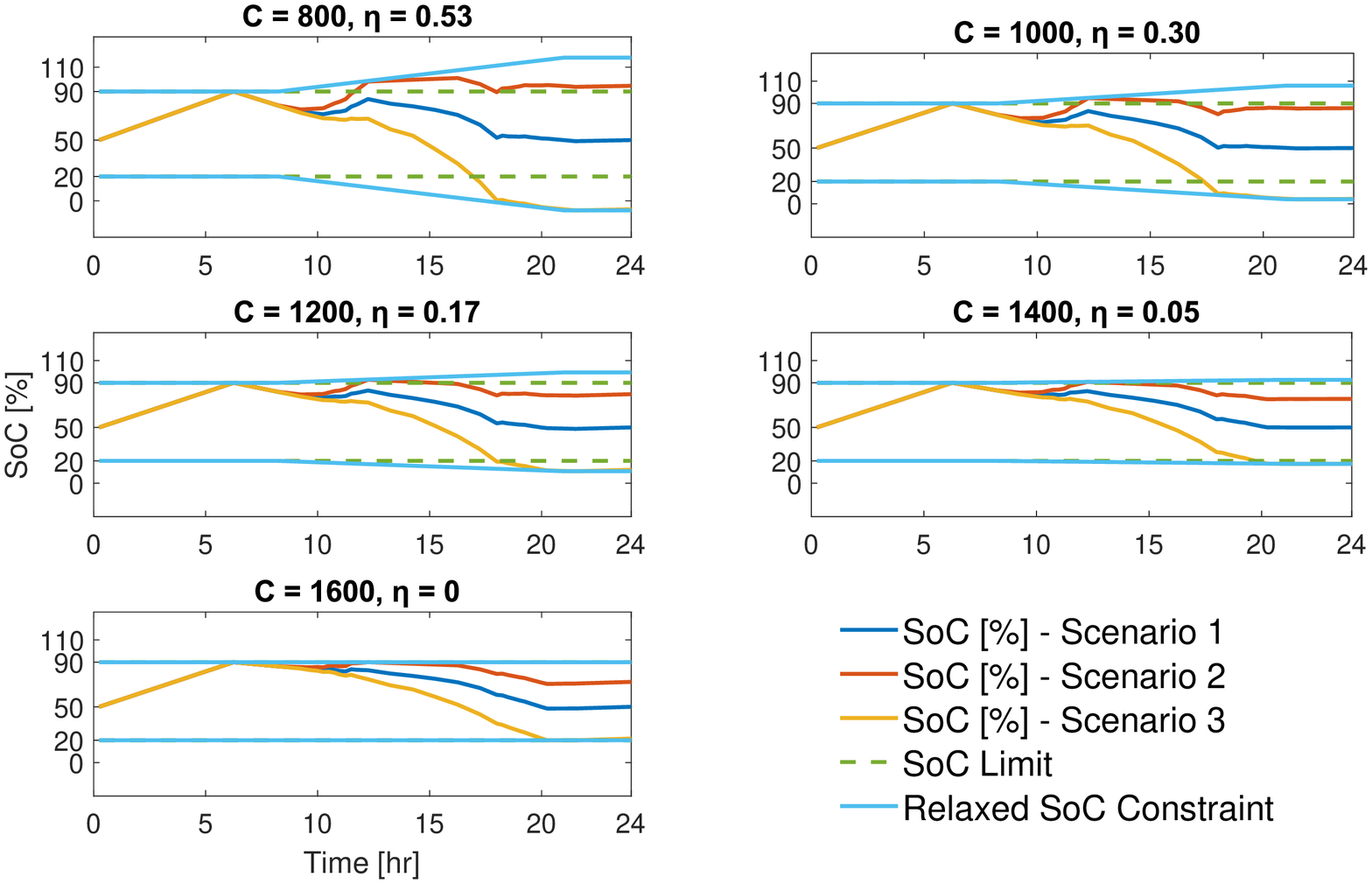}
    \caption{Effect of storage size on scheduling with uncertainty. With increasing storage size, $\eta^*$ decreases meaning that relaxed constraints come closer to hard SoC constraints.}
    \label{SoC_min_eta}
\end{figure}


We seek to compute an economic schedule for inverter power that can be implemented regardless of the realization of solar profile within the estimated boundaries. An attempt to solve the problem with the given solar bounds and hard constraints of equation (\ref{SoC_constraint1}) on all solar scenarios reveals that no feasible solution exists. Even if such solution existed, it would be highly conservative due to the requirement that all possible scenarios should stay within bounds even for time intervals far from the current interval for which accurate predictions do not exist. Since we assume no knowledge of uncertainty distribution except its upper and lower bounds, CCP is not a good solution fit for our problem.

Next we investigate microgrid scheduling problem over 15-min intervals using the proposed algorithm. Algorithm 1 is implemented on a system with parameters 
$C=1$ MWh, 
$C_0=500$ kWh,
$SoC_{min}=20 \% $,
$SoC_{max}=90 \%$,
$P_{min,max}=-(+)250$ kW,
$\eta_0=1$, and
$\epsilon=0.01$. Gurobi commercial solver is used for solving the quadratic programming problem at each step. The resulting power profile as well as battery charge levels are presented. Figures \ref{Power} and \ref{SoC} show the result of scheduling at time $\tau=1$. Maximum power flow at PCC over the entire 24-hr interval is remarkably reduced by flattening PCC power profile. The inverter power profile has also low volatility in this case which makes it robust against unmodelled prediction errors. Evolution of battery charge level within the soft SoC constraints under three scenarios of solar generation is indicated in Figure \ref{SoC}. 

\begin{figure}[thpb]
    \centering
     \includegraphics[width=9.0cm]{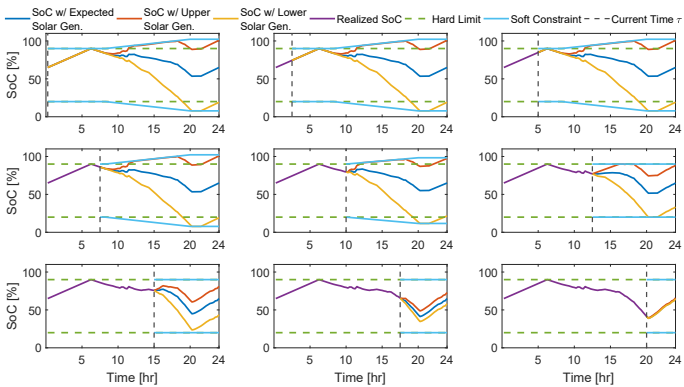}
    \caption{Evolution of battery SoC over the 24-hr interval shown at 9 steps during the day. The dashed line shows the current step of optimization and the purple line shows the actual realization of battery charge level for past times. While the results of optimization at earlier steps show apparent SoC limit violation at times far from the current step, no actual SoC violation occurs as time proceeds as the optimization is updated at every MPC step.}
    \label{MPC}
\end{figure}

To investigate the effect of storage size while solar uncertainty exists, Figure \ref{SoC_min_eta} shows economic scheduling results at the start of the 24-hr horizon. For microgrids with different storage sizes, the minimum viable $\eta~(\eta^*)$ for solution feasibility is obtained for each case. For smaller storage sizes (800, 1000, 1200, 1400), $\eta^*$ is greater than zero meaning that no feasible solution would exist had the soft constraints not replaced the hard ones. For storage of size 1600, the problem is solvable with $\eta=0$ or equivalently hard SoC constraints.


To illustrate the evolution of microgrid charge level over the 24-hr interval, Figure \ref{MPC} shows the evolution of battery charge level as the result of scheduling over the 24-hr interval. It is seen that as time proceeds and updated solar predictions become available, soft SoC constraints become tighter and no violation of hard SoC limits is observed at the end of the 24-hr interval.

\section{CONCLUSIONS}

A robust microgrid scheduling algorithm is proposed to optimize power exchange between microgrid and the main grid with uncertain solar prediction. To implement the algorithm, it suffices to have upper and lower bounds on solar generation to describe such uncertainty. By relaxing the original constraint on the charge level of the battery, optimal solutions are sought in a larger space of possible battery charge levels. The problem is formulated as a quadratic program and is solved at 15-min steps over a 24-hr interval and updated prediction data is used for each step. The model predictive formulation of the problem ensures that the apparent hard constraint violation does not lead to charge level requirement violation. The results indicate scheduling profiles that are in agreement with the defined cost function and follow the expected requirements under different uncertainty scenarios. Also, the effect of storage size on handling the uncertainty is investigated.


\balance
\bibliographystyle{IEEEtran}
\bibliography{IEEEabrv,mybib}

\end{document}